\documentclass[aps,twocolumn,showpacs,amssymb,prl]{revtex4-1}
\date{\today}
\usepackage{graphicx}
\usepackage{color}

\usepackage{array}
\usepackage{graphicx}
\usepackage{amssymb}
\usepackage{longtable}

\usepackage{bm}
\begin{document}

\title{Novel Monte Carlo representation for shell model in the complex energy plane}
\author{Zhen-Xiang Xu}
\thanks{Present address: School of Physics, Peking University, Beijing 100871, China}
\affiliation{KTH (Royal Institute of Technology), Alba Nova University Center,
SE-10691 Stockholm, Sweden}
\author{Chong Qi}
\thanks{Email: chongq@kth.se}
\affiliation{KTH (Royal Institute of Technology), Alba Nova University Center,
SE-10691 Stockholm, Sweden}\date{\today}

\begin{abstract}
A Monte Carlo method is presented to evaluate quantum states with many particles moving in the continuum. The scattering state is generated at each time by a Monte Carlo random sampling algorithm. The same calculation are repeated until the average energies of all calculations converge. For systems with one and a few particles, our calculations show that the exact solution can be approached with around only one hundred iterations. As examples, we apply the approach to study the structure of neutron-rich oxygen isotopes.
\end{abstract}

\maketitle
The configuration interaction approach plays an decisive role in the description of quantum many-body systems including quantum chemistry, atomic and molecular physics, condensed matter physics and nuclear physics. In nuclear physics it is more known as the interacting shell model in which the wave function is constructed as a linear expansion of all possible anti-symmetric Slater determinants within a given model space. The model space is usually defined by taking into account the single-particle orbitals within one or a few major shells. However, the size of the
model space increases exponentially with the number of valence
nucleons and orbitals, which soon becomes much larger than those treated by
conventional diagonalization techniques \cite{Lei12,Shi12}.

Another challenge is that the shell model practically describe the nucleus as a closed system since the continuum effect is not considered explicitly \cite{Mah69}. Among the attempts to tackle this problem we mention the shell model in the complex energy plane \cite{Betan02,Mic02,Mic09}
which spans the Berggren space \cite{Ber68}. This is made possible by the introduction of the so-called Berggren representation where the continuum is represented by an ensemble of discretized scattering states on a chosen contour \cite{Lio96}. The Berggren representation has been successfully applied to study the structure of light He, Li and O isotopes (see, e.g., Refs. \cite{Mic09,Rot06,Xu11}) and is also incorporated into the coupled-cluster method \cite{Hag12}. It is to be emphasized that the calculation within the Berggren representation is even more challenging than the traditional configuration interaction approach since a large amount of continuum states have to be included  \cite{Lio96}, which increase dramatically the configuration space, as well as make it non-Hermitian and complex \cite{Rot06,Xu11}. Moreover, it is difficult to identify the physically meaningful state from such calculations since they induce many states which can lie at very low energies but are immersed in the continuum.

In this letter we will develop an alternative Monte Carlo method to perform calculations in the continuum.
Our basic assumption is that resonances can be
described in terms of states lying in the complex energy plane. These complex states
correspond to solutions of the Schr\"odinger equation with outgoing boundary
conditions.
The eigenstates thus obtained can be used to
express the Dirac $\delta$-function as \cite{Ber68}
\begin{equation}\label{eq:delb}
\delta(r-r')=\sum_n w_n(r) w_n(r') + \int_{L^+} d\varepsilon u(r,\varepsilon) u(r',\varepsilon),
\end{equation}
where the sum runs over all the bound states and complex
states
(poles) which lie between the real energy axis and the integration contour
$L^+$.  The wave function of a state $n$ in these discrete set is
$w_n(r)$, and $u(r,\varepsilon)$ is the scattering function at energy $\varepsilon$.
The Berggren representation is obtained by discretizing the integral in Eq. (\ref{eq:delb}) \cite{Lio96}. It includes all the bound, antibound, resonance and a large number of discretized scattering states.
In this work, instead of including many scattering basis states, we will use only one or a few samples of them at a given time. The same calculation will be repeated until the average energies corresponding to all those calculations convergence to a certain value. We call this the Monte Carlo representation since the scattering state is generated at each step by the Monte Carlo random sampling algorithm. With this procedure the dimension of the basis, which can otherwise be many orders of magnitude larger than the one proposed here, is relatively easy to handle. Another advantage is that the physically meaningful state can be easily identified due to  the large removal of states mainly composed of continuum.
Moreover, the model we proposed can be straightforwardly incorporated into the variety of Monte Carlo truncation schemes of configuration interaction approaches \cite{Shi12,She12,Bon13} by generalizing them to the complex energy plane.

As an illustration, we firstly apply the method to evaluate the one-particle state in the complex energy plane for which exact solution exists.
We use a realistic single-particle Hamiltonian $H_0$ to generate several sets of orthonormal basis states. Within each set, we diagonalize a certain target Hamiltonian $H$ to evaluate its eigenvalues and eigenvectors. The exact solution is derived by solving directly the Schr\"odinger equation with proper boundary
conditions.
As in Ref. \cite{Lio96}, we use a Woods-Saxon potential corresponding to the nucleus $^{208}$Pb. We concentrate on two of its $h_{11/2}$ eigenstates: A bound state at $-14.960$ MeV and a resonance at $(2.251,-0.026)$ MeV. The target Hamiltonian $H$, which has a different potential depth than $H_0$, has one bound state at $-12.526$ MeV and one resonance at $(4.322,-0.321)$ MeV.

 To evaluate these states within the framework of complex shell model \cite{Betan02,Mic02,Mic09,Lio96},
we choose two contours in the complex energy plane with different cutoff energies. The nodes of each contour are as follows (in MeV):
Contour 1, $(0,0)\rightarrow(0,-5)\rightarrow(20,0)\rightarrow(30,0)$;
Contour 2, $(0,0)\rightarrow(0,-5)\rightarrow(20,0)\rightarrow(30,0)\rightarrow(100,0)$.
For each contour, we select the same number $n$ of discretized scattering states $u(r,\varepsilon_p)$ on different segments. The weights $h_p$ are calculated according to Gaussian quadrature method \cite{Lio96}.  We diagonalized the target Hamiltonian $H$ within these two bases. The results correspond to different $n$ are given in Table \ref{old1p}. it is seen that, when $n$ is small, the difference between the calculated eigenvalue and the exact result is quite large, especially for the resonance state. Moreover, the converged eigenvalues for calculations within contour 1 differ from the exact values by about 2 keV. It means that the scattering states with the energy larger than 30 MeV still have noticeable contribution to the wave function. Our calculations show that scattering states with energy higher than 100 MeV have negligible influence.

\begin{table}
\begin{center}
\caption{The calculated one-particle eigenvalues within different contours as a function of the number $n$ of scattering states on each segment. The energies are in units of MeV. 
}\label{old1p}
\begin{ruledtabular}
\begin{tabular}{ccrcccccc}
      Basis          &&   n   &&  Bound State  && Resonance  &\cr
\hline
            && 2     &&   $-12.519$     &&$(4.442,-0.209)$ &\cr
            && 4     &&   $-12.525$     &&$(4.314,-0.372)$ &\cr
            && 8     &&   $-12.525$     &&$(4.343,-0.340)$ &\cr
Contour 1   && 16    &&   $-12.524$     &&$(4.329,-0.321)$ &\cr
            && 32    &&   $-12.524$     &&$(4.324,-0.322)$ &\cr
            && 64    &&   $-12.524$     &&$(4.324,-0.322)$ &\cr
            && 128   &&   $-12.524$     &&$(4.324,-0.322)$ &\cr
\hline
            && 2     &&   $-12.521$     &&$(4.429,-0.209)$ &\cr
            && 4     &&   $-12.527$     &&$(4.312,-0.372)$ &\cr
            && 8     &&   $-12.526$     &&$(4.341,-0.339)$ &\cr
Contour 2   && 16    &&   $-12.526$     &&$(4.327,-0.321)$ &\cr
            && 32    &&   $-12.526$     &&$(4.322,-0.321)$ &\cr
            && 64    &&   $-12.526$     &&$(4.322,-0.321)$ &\cr
            && 128   &&   $-12.526$     &&$(4.322,-0.321)$ &\cr
\hline
Exact       &&       &&   $-12.526$     &&$(4.322,-0.321)$ &\cr
\end{tabular}
\end{ruledtabular}
\end{center}
\end{table}

\begin{figure}
\includegraphics[width=0.4\textwidth]{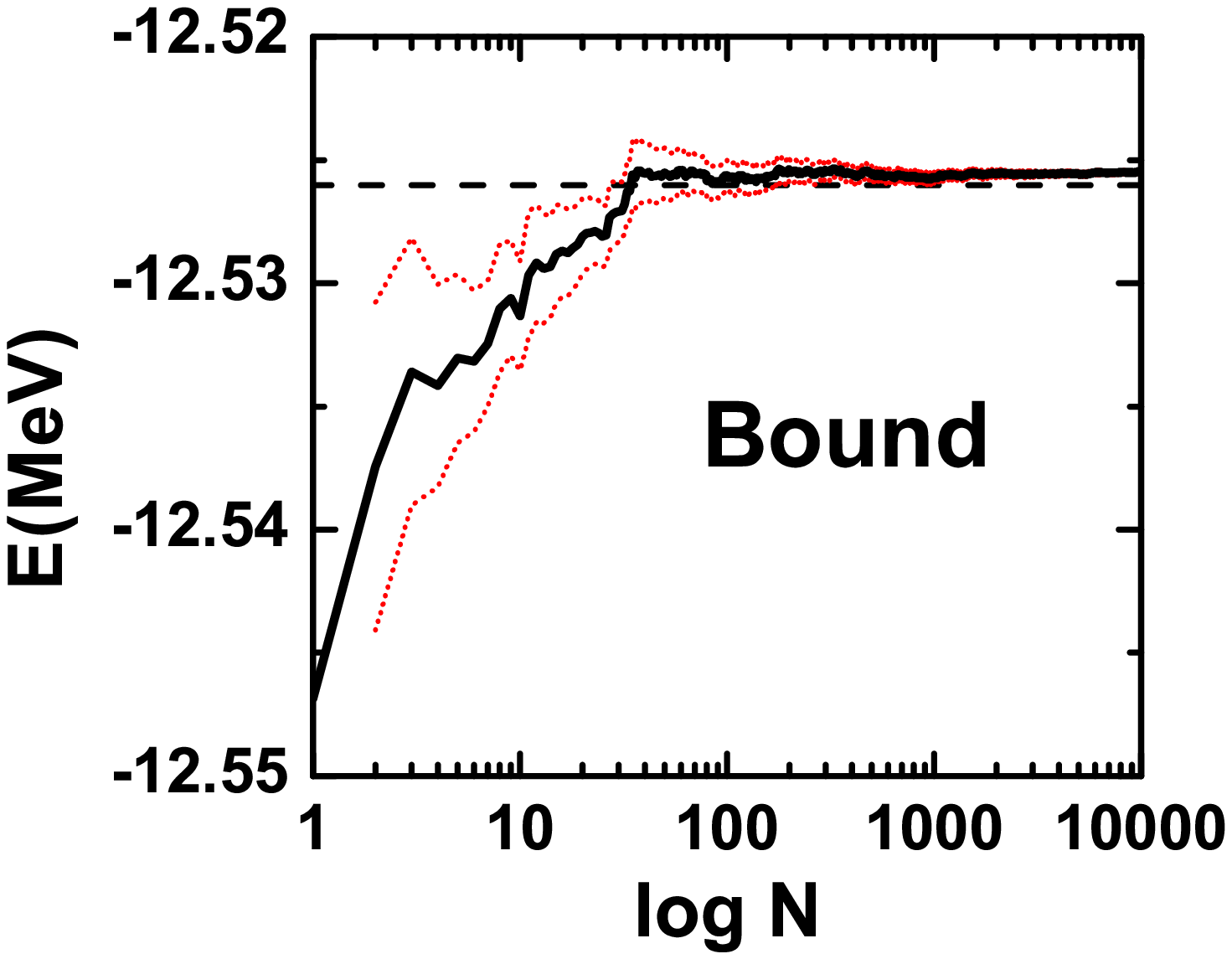}\\
\includegraphics[width=0.4\textwidth]{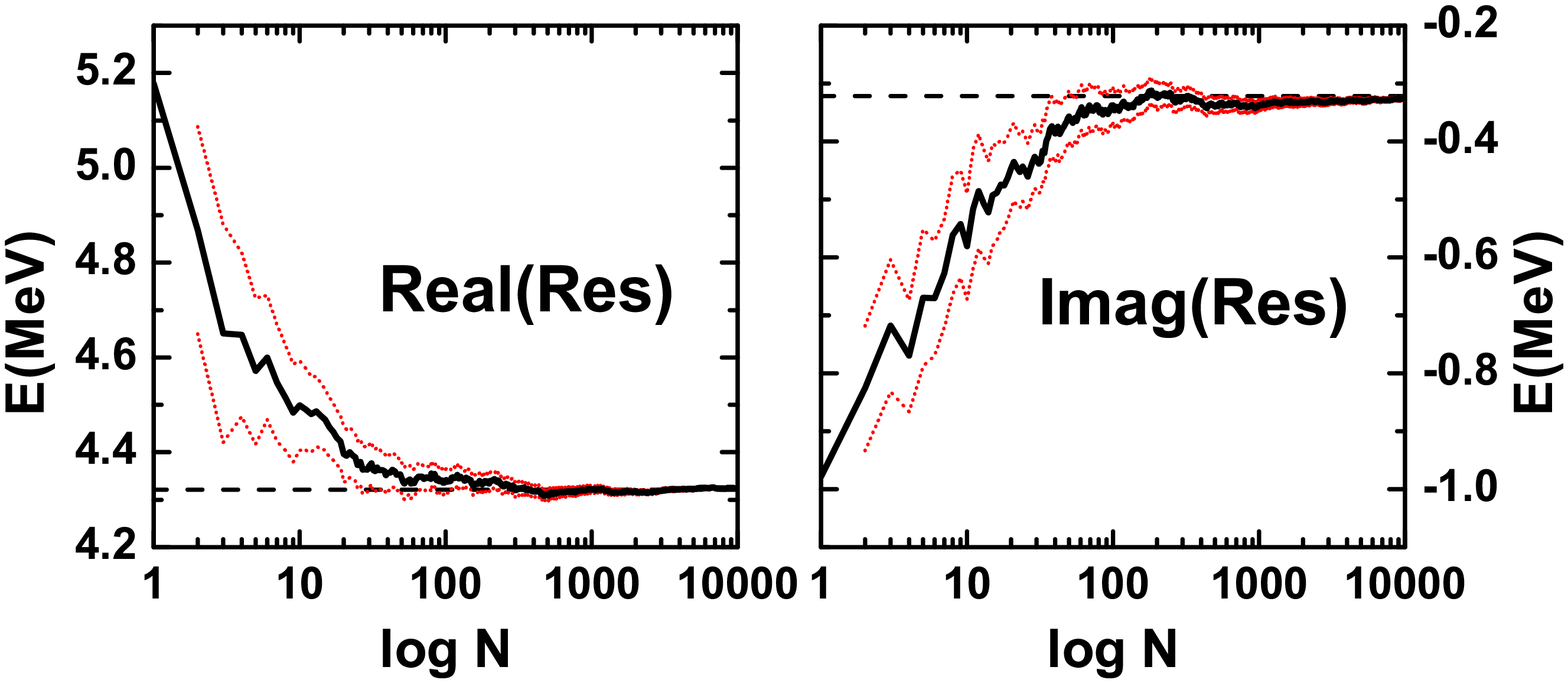}\\
\caption{Upper: Convergence of the Monte Carlo calculation for the  one-particle bound state. The black solid line denotes the calculated average value while the red lines indicate the estimated error. The straight dashed line corresponds to the exact solution. Lower: Same as the upper panel but for the Resonance. The left and right plots show the real and imaginary parts of the eigenvalue, respectively.}\label{MCb}
\end{figure}

Instead of summing many discretized scattering states over the contour, we will use a Monte Carlo method to approach the integral. We approximate the integral in Eq. (1) as
\begin{equation}
\int_{L^+} d\varepsilon u(r,\varepsilon) u(r',\varepsilon) \approx \frac{1}{N}\sum_{i=1}^N\sum_{p} v_pu(r,\varepsilon_{pi}) u(r',\varepsilon_{pi}),
\end{equation}
where $p$ runs over the different segments of $L^+$ and $v_p$ denotes the length of the corresponding segment.
In practice, we will choose three segments for the contour $(0,0)\rightarrow(0,-5)\rightarrow(20,0)\rightarrow(100,0)$ and randomly pick up one scattering sample on each segment at each time $i$. Now our shell-model basis consists only one bound state, one resonance and three random scattering states. In the other word, we consider, at a discrete time, one scattering state in each step a `single' quantum state which is the superposition of several ones, $|\phi(r)_i\rangle=\sum_p\sqrt{v_p/\cal{N}}|u(r,\varepsilon_{pi})\rangle$ where $\cal{N}$ denotes the normalization factor \cite{Des10}. Within this basis, for which the dimension is much smaller than that of complex shell model, we diagonalize at each time $i$ the target Hamiltonian $H$ as before. The same calculation is repeated $N$ times. The final total energy $E$ of a given state is given as the average value of all the corresponding eigenvalue $E_i$ we got in previous calculation. The physically meaningful states one get at each time can be easily identified, since they are dominated by the pole configurations. The convergence of our calculations for the bound and resonant states are shown in Fig. \ref{MCb}. The converged results are $-12.526$ MeV for the bound state and $(4.324, -0.328)$ MeV for the resonance after 50,000 individual runs. The systematic error for our calculation of the bound state energy is less than 1 keV. The error for the resonance state is only about 3 keV.

Now we apply our Monte Carlo representation to evaluate two-particle systems. We take the nucleus $^{26}$O as an example, which has two neutrons outside the closed shell $^{24}$O \cite{Stan04,Ele07,Hof09,Kan09,Hof08} and is expected to be unbound \cite{Lun12,Cae12,Koh13}. The nucleus $^{25}$O is also detected to be particle instable and decays under the emission of one neutron~\cite{Hof08}. The single particle states can be well  reproduced by using the Woods-Saxon potential \cite{Tsu09,Xu12}. The parameters we take correspond to $r_0=1.285$ fm, $V_{so}=16.331$ MeV, $r_{so}=1.146$ fm and $a=0.691$ fm \cite{Xu12}. The depth $V_0$ is slightly adjusted to reproduce the experimental energy of the $d_{3/2}$ orbital at $770$ KeV \cite{Hof08}.

 In $^{26}$O the lowest valence shells are $d_{3/2}$ and $f_{7/2}$. The ground state of $^{26}$O is measure to be at $0.150$ MeV \cite{Lun12}. For comparison, we first use Complex Shell Model Method (CXSM) to calculate the $0^+$ states of $^{26}$O. For the body matrix elements we take the simple separable force \cite{Betan02,Xu11}. The strength $G$ of the separable interaction is determined by fitting to the ground state energy. We chose contour $(0,0)\rightarrow(0,-3)\rightarrow(10,-5)\rightarrow(40,-5)\rightarrow(100,0)$ MeV for $d_{3/2}$, and contour $(0,0)\rightarrow(0,-5)\rightarrow(25,-5)\rightarrow(100,0)$ MeV for $f_{7/2}$. We take 15 scattering states on each segment, which means one has 105 scattering states in total. The results of the calculation are shown in Table \ref{com2p}. we also did the calculation without including the continuum states. In this case the model space only contains two resonances $d_{3/2}$ and $f_{7/2}$. As is shown in Table \ref{com2p}, the ground state thus obtained has a positive imaginary part of the energy, which is not physical, due to the non-completeness of the basis.

\begin{table}[b]
\begin{center}
\caption{The energies of the $0^+$ states in $^{26}$O given by the shell-model calculation without continuum, complex shell model (CXSM) and our Monte Carlo calculations MC-I and MC-II. The energies are in units of MeV.}\label{com2p}
\begin{ruledtabular}
\begin{tabular}{ccccccccc}
                    &&  $^{26}$O$(0^+_1)$  &&  $^{26}$O$(0^+_2)$  &\cr
\hline
No Continuum   &&  $0.150+i\,0.743$   &&  $6.330-i\,0.448$   &\cr
    CXSM           &&  $0.150-i\,0.000$   &&  $6.780-i\,0.515$   &\cr
MC-I &&  $0.150-i\,0.000$   &&  $6.779-i\,0.517$   &\cr
MC-II &&  $0.149-i\,0.060$   &&  $6.780-i\,0.519$   &\cr
\end{tabular}
\end{ruledtabular}
\end{center}
\end{table}

\begin{figure}
\centering
\includegraphics[width=0.48\textwidth]{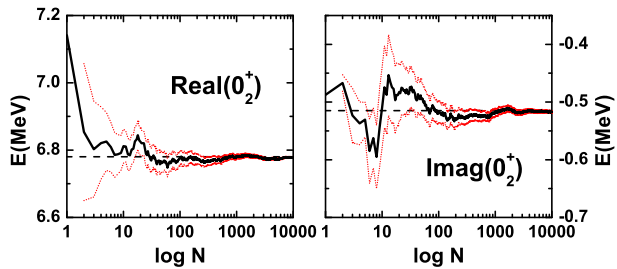}\\[.5em]
\caption{Same as Fig. \ref{MCb} but for the two-particle states $0^+_2$, in $^{26}$O, derived from the MC-I calculation. The left and right panels show the real and imaginary parts of the eigenvalue, respectively. The straight dashed lines correspond to the CXSM calculation.}\label{mc2pfgs}
\end{figure}

\begin{figure}
\centering
\includegraphics[width=0.48\textwidth]{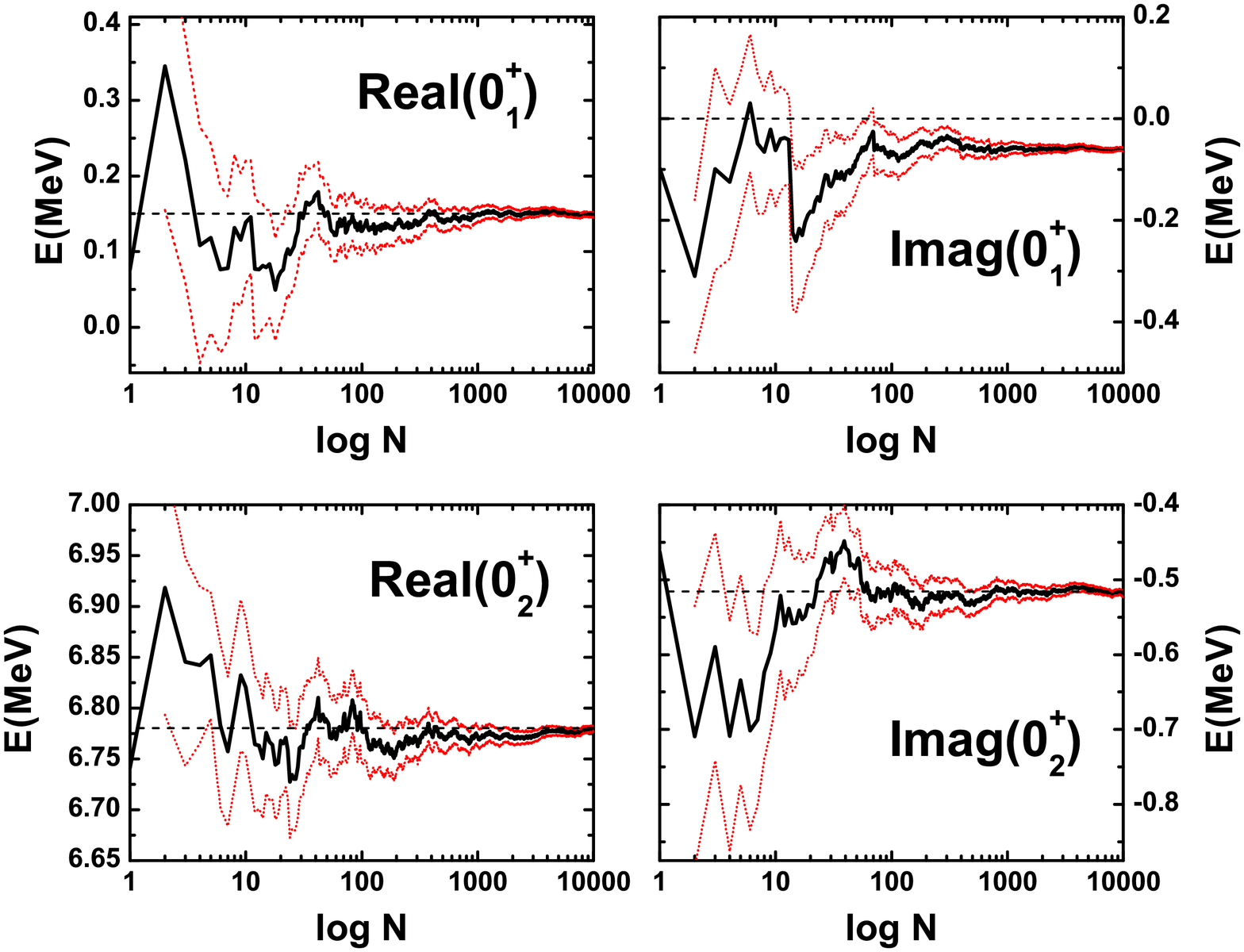}\\[.5em]
\caption{Same as Fig. \ref{mc2pfgs} but for the two-particle ground state (upper penal) and the $0^+_2$ state (lower penal), in $^{26}$O derived from the MC-II calculation.}\label{mc2pfg}
\end{figure}

Now we proceed to do the same calculations by applying the Monte Carlo method. We use the same contour as in the CXSM calculation. Two kinds of calculations are done. Firstly, the energies of $0^+$ states in $^{26}$O are calculated by fixing the strength $G$ in each run to reproduce the energy of the ground state (denoted as MC-I). The convergence of our calculations for the excited state $0^+_2$ is plotted in Fig. \ref{mc2pfgs}. We repeat the procedure for 10,000 times and the results shown in Table \ref{com2p} are the converged average of all individual runs. We can see that the accuracy of our Monte Carlo calculation is quite satisfactory. The difference between the Monte Carlo and CXSM calculations is below $2$ keV.

For the second step, we redo the Monte Carlo calculation by taking for the strength $G$ to the same value as in CXSM (MC-II). The convergence of our calculations for the ground state and the first excited $0^+$ state are plotted in Fig. \ref{mc2pfg}. We repeat the procedure for 10,000 times and the converged average are also shown in Table \ref{com2p}. The accuracy of the Monte Carlo calculation is also good. The difference between the Monte Carlo and CXSM calculations is below $1$ keV, except the imaginary part of the ground state has a difference of $60$ keV. The reason for the relatively large difference might be that the ground state as well as the single-particle  $d_{3/2}$ resonance are close to the origin. Hence the correlation between the resonance and the continuum around the origin is pretty strong. This discrepancy may be fixed by taking more scattering states close to origin in the Monte Carlo calculations.

\begin{figure}
\centering
\includegraphics[width=0.48\textwidth]{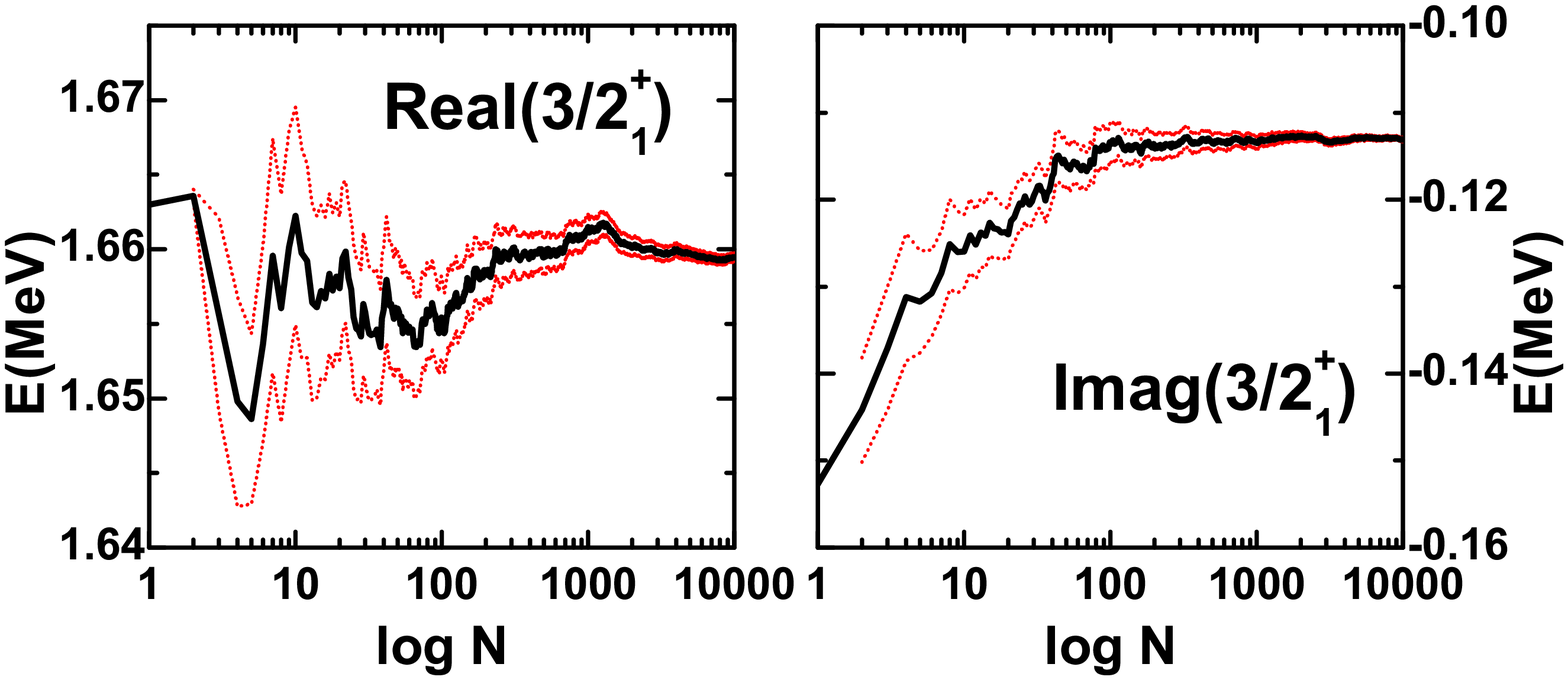}\\
\includegraphics[width=0.48\textwidth]{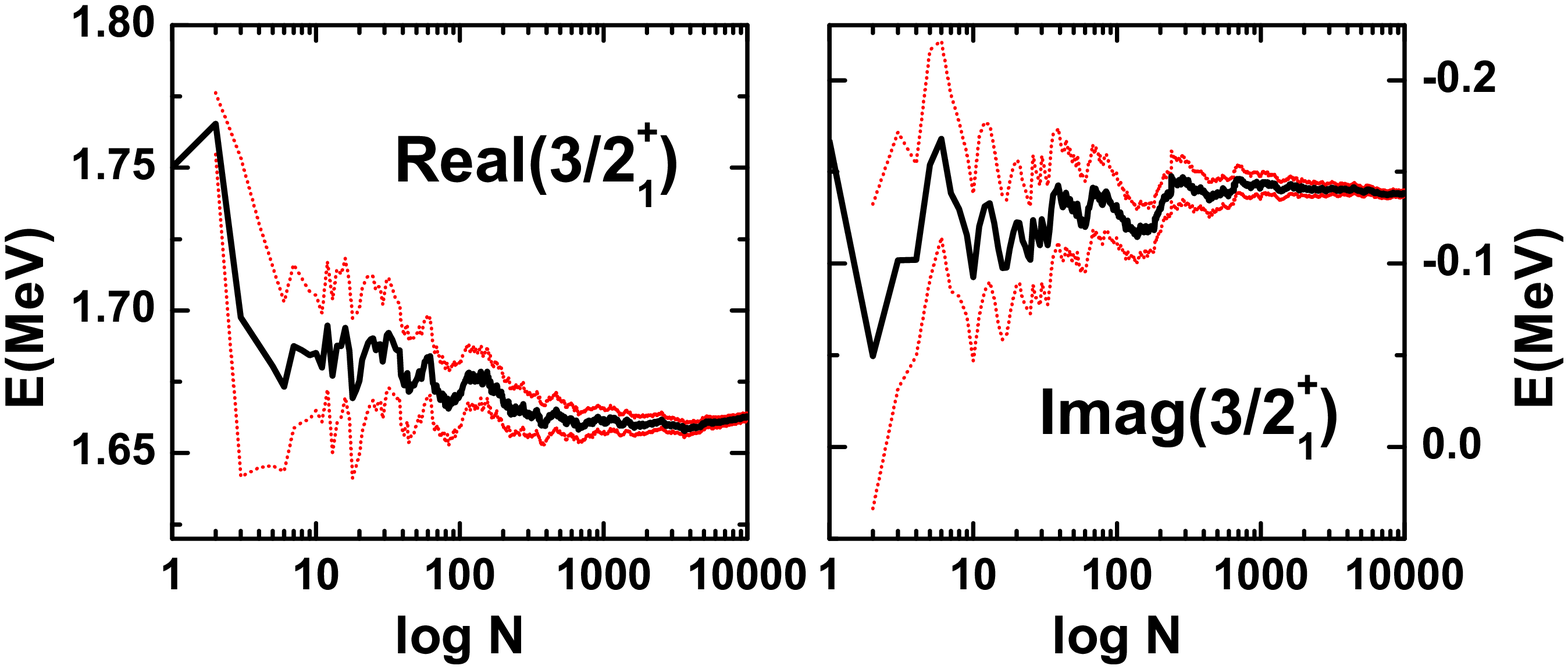}\\[.5em]
\caption{Same as Fig. \ref{MCb} but for the three-particle ground states $3/2^+_1$, in $^{27}$O, derived from the MC-I (upper) and MC-II (lower) calculations.}\label{mc3pfg}
\end{figure}

We now apply our Monte Carlo representation to calculate the three-particle system $^{27}$O. The three-particle bases are constructed by applying the same approach as Ref. \cite{Xu11}. Only the pairing interaction is taken into account for simplicity. The ground state energy is calculated to be $1.659-i0.113$ and $1.663-i0.138$ MeV for calculations with the MC-I and MC-II approaches, respectively. The convergence of the ground state with different approaches are given in Fig. \ref{mc3pfg}. The first excited state is calculated to be $7/2^-$. The energies are calculated to be $4.605-i0.556$ MeV (MC-I) and $4.604-i0.614$ MeV (MC-II).

In principle, the effective interaction should be renormalized differently for calculations within different model spaces (see, e.g., Ref. \cite{Jensen}). However, our calculations shown above tend to suggest that one can safely take the same interaction for calculations with the full CXSM and Monte Carlo calculations even though the number of single-particle states is highly restricted in the latter case.

In full configuration interaction calculations, we do not distinguish between the discrete states and the `discretized' scattering state. Thus the total $M$-scheme dimension will be significantly larger when the continuum is taken into account. For CXSM calculations on $^{26}$O, the total $M$-scheme dimension is ${m \choose n}=66$ and 186966 for our calculations without and with the continuum states, respectively, where $m$ ($n$) denote the numbers of $M$-scheme orbitals (particles). In the latter scheme, it is impossible to treat systems with more than three particles on modern computers. It should be mentioned that many of those large amount of configurations, which are mostly composed of particles in the continuum, are not physically meaningful and play a relatively minor role in the final wave function. This redundancy is mostly removed in our Monte Carlo calculation, for which is the corresponding dimension is ${m \choose n}=1326$, even though one may argue that it is not a `full' representation. In that sense, the Monte Carlo method we propose is consistent with the variety of importance truncation algorithms employed in traditional configuration interaction approaches.

In summary, a Monte Carlo method is presented to evaluate the effect of the continuum within the configuration interaction approach. The scattering state on the contour is generated at each time by a Monte Carlo random sampling algorithm. We repeat the same calculation until the average energies of all calculations converge. We show that, for systems with one and two particles, the exact solution can be approached with only around one hundred iterations. With this procedure the dimension of the basis is much smaller than that of the traditional complex shell model. Moreover, the physically meaningful state can be identified in a more  straightforward way. The approach is employed to evaluate the energies and the structure of the heavy unbound oxygen isotopes $^{26,27}$O.

We thank R. Liotta, R. Wyss and F.R. Xu for stimulating discussions and all their support. We also acknowledge the discussions with M. P{\l}oszajczak and R. Id Betan. This work is supported by the
Swedish Research Council (VR) under grant Nos. 621-2010-4723 and 621-2012-3805. Z.X. is supported in part by the China Scholarship Council under grant No. 2008601032, the National Key Basic Research Program of China under Grant No. 2013CB834400, and the National Natural Science Foundation of China under Grant No. 11235001.

\end{document}